\documentclass[12pt,a4paper]{article}
\usepackage{type1cm,cite,amsmath,graphicx,indentfirst}
\usepackage[psamsfonts]{amssymb}
\usepackage{color}
\usepackage{tikz}
\numberwithin{equation}{section}

\newtheorem{thm}[subsection]{Theorem}

\newcommand {\g}{\mathfrak g}

\newcommand {\del}{\partial}

\newcommand{\str}{\text{str}}

\def\cV{{\mathcal V}}
\def\ra{\rightarrow}

\newcommand{\al}{\alpha} \newcommand{\ga}{\gamma}
 \newcommand{\be}{\beta}
\newcommand{\ka}{\kappa} \newcommand{\de}{\delta}
\newcommand{\ep}{\epsilon}

\newcommand{\affine}[1]{\widehat{#1}}
\newcommand{\alg}[1]{\mathfrak{#1}}

\newcommand{\SLA}[2]{\alg{#1} \left( #2 \right)}
\newcommand{\SLSA}[3]{\alg{#1} \left( #2 \middle\vert #3 \right)}
\newcommand{\AKMA}[2]{\affine{\alg{#1}} \left( #2 \right)}
\newcommand{\AKMSA}[3]{\affine{\alg{#1}} \left( #2 \middle\vert #3 \right)}

\begin{document}
\begin{titlepage}

 \renewcommand{\thefootnote}{\fnsymbol{footnote}}
\begin{flushright}
 \begin{tabular}{l}
 arXiv:1112.xxxx\\ 
 \today 
 \end{tabular}
\end{flushright}

 \vfill
 \begin{center}


\noindent{\large \textbf{Fermionic Coset, Critical Level $\mathcal{W}^{(2)}_4$-Algebra and Higher Spins }}\\
\vspace{1.5cm}

\noindent{ Thomas Creutzig,$^{a}$\footnote{E-mail: tcreutzig@mathematik.tu-darmstadt.de} Peng Gao$^b$\footnote{E-mail: pgao@scgp.stonybrook.edu} and Andrew R. Linshaw$^{c}$\footnote{E-mail: linshaw@brandeis.edu}}
\bigskip

 \vskip .6 truecm
\centerline{\it $^a$Fachbereich Mathematik,
Technische Universit\"{a}t Darmstadt,}
\centerline{\it Schlo\ss gartenstr. 7, 64289 Darmstadt, Germany}
\medskip
\centerline{\it $^b$ Simons Center for Geometry and Physics, Stony Brook University,}
\centerline{\it  Stony Brook, NY 11794-3840, USA}
\medskip
\centerline{\it $^c$ Department of Mathematics, Brandeis University
} \centerline{\it Waltham, MA 02453, USA
}
 \vskip .4 truecm

 \end{center}

 \vfill
\vskip 0.5 truecm

\begin{abstract}

The fermionic coset is a limit of the pure spinor formulation of the AdS$_5\times$S$^5$ sigma model as well as a limit of a nonlinear
topological A-model, introduced by Berkovits.
We study the latter, especially its symmetries, and map them to higher spin algebras.

We show the following.
The linear A-model possesses affine $\AKMSA{pgl}{4}{4}_0$ symmetry at critical level and
its $\AKMSA{psl}{4}{4}_0$ current-current perturbation is the nonlinear model.  
We find that the perturbation preserves $\mathcal{W}^{(2)}_4$-algebra symmetry at critical level.
There is a topological algebra associated to $\AKMSA{pgl}{4}{4}_0$ with the properties that the 
perturbation is BRST-exact. Further, the BRST-cohomology contains world-sheet supersymmetric symplectic fermions
and the non-trivial generators of the $\mathcal{W}^{(2)}_4$-algebra.
The Zhu functor maps the linear model to a higher spin theory. We analyze its $\SLSA{psl}{4}{4}$ action and find finite dimensional short multiplets.
\end{abstract}
\vfill
\vskip 0.5 truecm

\setcounter{footnote}{0}
\renewcommand{\thefootnote}{\arabic{footnote}}
\end{titlepage}

\newpage


\section{Introduction}

The AdS/CFT correspondence \cite{Maldacena:1997re} proposed a duality between type IIB superstring theory in AdS$_5\times$S$^5$ and ${\cal N}=4$ supersymmetric Yang-Mills theory in four dimensions, and is believed to apply much more generally in the context of strongly coupled field theories. However,  extensions of the original proposal mostly do not shed new light on the origin of this deep idea or take us away from the weakly curved AdS space regime. 
To understand the full content of AdS/CFT correspondence, it is crucial to further reveal the connection between weakly coupled gauge theories and string theory in AdS in the small radius case.

This is a difficult problem as in the small radius limit the string theory world-sheet theory is expected to be strongly interacting and hard to understand.
It was proposed that the bulk theory should contain infinitely many fields of high spin corresponding to conserved currents of the free boundary theory \cite{Konstein:2000bi,Sundborg,EWtalk,Shaynkman:2001ip,Sezgin:2001zs,Vasiliev:2001zy,Polyakov:2001af,Mikhailov:2002bp,Sezgin:2002rt}. From the string theory point of view, this arises from a tensionless limit of the world-sheet theory \cite{Sundborg,Isberg:1992ia,Segal:2002gd,Bianchi:2003wx,Lindstrom:2003mg,Bonelli:2003zu,Sagnotti:2003qa,Bakas:2004jq}. In fact, Polyakov pointed out quite early on in \cite{Polyakov:2001af} a possible mechanism by which the strongly coupled world-sheet sigma model could preserve the integer space-time conformal dimensions of certain high derivative operators. Based on large $N$ factorization arguments put forward in \cite{Sundborg,EWtalk}, Mikhailov \cite{Mikhailov:2002bp} argued that the (free) classical bulk theory can be reconstructed nearly uniquely in the strict $N\rightarrow \infty$ limit from the free $SU(N)$ SYM theory on the boundary.  

 Despite the early progress, the fully interacting higher spin bulk theory is not known for AdS$_5$ (see however \cite{Vasiliev:2001wa}).  In addition, in \cite{Klebanov:2002ja} it was pointed out that the interacting higher spin theories of Fradkin-Vasiliev type \cite{Fradkin:1986qy} may not contain enough bulk fields to be dual to large $N$ field theories with elementary fields in the adjoint representation, such as ${\cal N}=4$ supersymmetric Yang-Mills theory. In fact, this motivated their proposal of a duality between Fradkin-Vasiliev type higher spin gauge theory and the critical $O(N)$-vector model with fields in the fundamental representation, which recently received substantial evidence from highly non-trivial bulk \cite{Giombi:2009wh} and boundary \cite{Douglas:2010rc} calculations. 
 
 In this paper, we follow a different approach, as first proposed by Berkovits in a series of papers \cite{Berkovits:2007zk,Berkovits:2007rj,Berkovits:2008qc,Berkovits:2008ga}. There, an exotic scaling limit of the world-sheet fields was taken in analogy with the usual flat space limit of AdS$_5\times$S$^5$, leading to a topological gauged linear sigma model with fermionic superfields, conjectured to describe the zero radius limit of the AdS$_5\times$S$^5$ superstring. In this limit, the superspace torsion values are different from that of any standard supergravity backgrounds. Moreover, the action is quadratic and can be considered as the large volume limit of a topological nonlinear sigma model. In \cite{Berkovits:2007rj} the ambitious goal of deriving AdS/CFT correspondence from the world-sheet theory was proposed, relying on formal similarities with the large $N$ Chern-Simons/topological A-model duality studied in \cite{Berkovits:2003pq}. 
 
 The nonlinear sigma model introduced by Berkovits is a fermionic coset model. We study this fermionic coset in some detail, especially in relation to the existence of higher spin symmetry algebra on the world-sheet. This is required if we want to identify the limit as the zero radius limit of  AdS$_5\times$S$^5$ superstring which is dual to the free limit of ${\cal N}=4$ supersymmetric Yang-Mills theory. 
 
 We have a few interesting findings in this paper. We show that the nonlinear A-model is a $\AKMSA{psl}{4}{4}_0$ current-current perturbation of the linear model which has an affine $\AKMSA{pgl}{4}{4}_0$ symmetry at critical level. The most interesting feature of the current-current perturbation is that it preserves a $\mathcal{W}^{(2)}_4$-algebra at critical level. This is a higher spin world-sheet symmetry algebra, as first classified by Zamolodchikov \cite{Zamolodchikov:1985wn}. The particular example of higher spin algebra we find is believed to be the Drinfeld-Sokolov reduction of $\AKMA{sl}{4}_{-4}$ for a non-principal embedding of $\SLA{sl}{2}$. It belongs to a class of such algebras  first studied by Feigin and Semikhatov \cite{Feigin:2004wb}. These algebras are closely related to affine Lie super algebras. Feigin and Semikhatov obtained them as cosets of $\AKMSA{sl}{n}{1}$, while they can also be realized at special levels from $\AKMSA{gl}{1}{1}$ \cite{CR}.
 
 Notice that in the nonlinear A-model there is a different BRST-operator from the one of the pure spinor AdS$_5\times$S$^5$ superstring, so it is not surprising that we find a non-trivial cohomology in the world-sheet closed string sector, different from the proposal in \cite{Berkovits:2007zk} where the bulk cohomology was assumed to be trivial. The cohomology contains world-sheet supersymmetric symplectic fermions and the non-trivial generators of the $\mathcal{W}^{(2)}_4$-algebra. 
Then using the machinery of Zhu's functor \cite{Zh}, we map the free world-sheet theory of the linear model to a higher spin super algebra extending $\SLSA{psl}{4}{4}$, and also the $\mathcal{W}^{(2)}_4$-algebra of the nonlinear A-model to a subalgebra  ${\cal Z}^{(2)}_4$ of the higher spin algebra.
 
In the present work, we make a connection to affine Lie super algebras. 
This is quite remarkable for the following reason.
Conformal field theories with affine Lie super algebra as symmetry are Lie super group Wess-Zumino-Witten models.
This class of theories is fairly well-understood 
\cite{Rozansky:1992rx,Schomerus:2005bf,Saleur:2006tf,Gotz:2006qp,Quella:2007hr,Creutzig:2007jy,Creutzig:2009fh,Hikida:2007sz,Creutzig:2009zz,Creutzig:2008an,Creutzig:2008ag,Creutzig:2008ek,Creutzig:2010zp,Creutzig:2011cu,Creutzig:2011qm}.
If the Killing form of the underlying Lie super algebra vanishes, then current-current perturbations
are truly marginal. Further they are under good control and physical quantities can be computed 
\cite{Quella:2007sg,Candu:2009ep,Mitev:2008yt,Konechny:2010nq,Creutzig:2010hr,Candu:2010yg,Ashok:2009xx,Benichou:2010rk}.
So in principle one can apply the results of these works to the nonlinear A-model.

This article is organized as follows. 
In section two, we explain how to rewrite the action of the nonlinear A-model of Berkovits. 
In section three, we find that the nonlinear A-model is a $\AKMSA{psl}{4}{4}$ current-current perturbation 
at critical level. In the large radius limit, the theory is a free theory with $\AKMSA{pgl}{4}{4}_0$ symmetry.
We then discuss symmetries of the theory. Most importantly, we find that the perturbation preserves the critical level $\mathcal{W}^{(2)}_4$-algebra.
In section four, we analyze the Zhu algebra of the large radius limit. This becomes a type of higher spin algebra extending $\SLSA{psl}{4}{4}$. We decompose the corresponding $\SLSA{psl}{4}{4}$ representations and find that they all correspond to short-multiplets. The appendix deals with some technical issues.

\section{The Sigma Model Action}

In \cite{Berkovits:2007zk} Nathan Berkovits introduced a new limit of the AdS$_5\times$S$^5$ sigma model
corresponding to the fermionic coset
\begin{equation}
\frac{\text{PSU}(2,2|4)}{\text{SU}(2,2)\times\text{SU}(4)}\, .
\end{equation}
This is a small radius limit. The sigma model in the pure spinor formulation
has a twisted $\mathcal N=2$ superconformal algebra. Since Berkovits 
considered the boundary theory with A-type boundary conditions he called this coset the linear A-model.
He then introduced a nonlinear A-model whose large radius limit is the linear A-model.

The purpose of this section is to write down the action of the nonlinear A-model.
We start with a construction of actions of chiral models, that is sigma models of cosets of group manifolds 
corresponding to an involutive automorphism.

\subsection{Chiral Model Action}

Let $G$ be a Lie group or supergroup. Let $\omega$ be an automorphism of $G$ of order two, such that the induced automorphism of the Lie algebra or Lie super algebra $\mathfrak g$ of $G$ preserves the invariant bilinear form. The subgroup fixed by $\omega$ we call $H$, and its Lie algebra $\mathfrak h$. We now explain the action of a sigma model for the coset $G\setminus H$. As this model only has a kinetic term, but no antisymmetric one, it is called a chiral model. It is to be of the form
\begin{equation}\label{eq:chiralmodel}
 S_{G\setminus H} \ = \ \int \str(g^{-1}\del g g^{-1}\bar\del g)_{\mathfrak g\setminus \mathfrak h}
\end{equation}
where the supertrace is restricted over the orthogonal complement of $\mathfrak h$ in $\mathfrak g$.  

The coset action is conveniently expressed in terms of the WZW model action.
Let $g$ be a map from the world-sheet to $G$. The WZW action is $S[g]=S_{\text{kin}}[g]+S_{\text{WZ}}[g]$, where 
the kinetic term of the Wess-Zumino-Witten model action is
\begin{equation}
	S_{\text{kin}}[g] \ = \ \frac{1}{2\pi}\int d\tau d\sigma\  \str( g^{-1}\del g\ g^{-1}\bar{\partial}g),
\end{equation}
and the Wess-Zumino term is
\begin{equation}
	S_{\text{WZ}}[g] \ = \ \frac{1}{6\pi}\int_B   d\tau d\sigma dt\
	\str( \tilde{g}^{-1}\del_t \tilde{g}\ \tilde{g}^{-1}\del \tilde{g}\ \tilde{g}^{-1}\bar\del \tilde{g}) 
	  \ , \vspace{2mm}
\end{equation}
where $\tilde g$ is an extension of $g$ to a 3-manifold $B$ with boundary the world-sheet as usual.
The Polyakov-Wiegmann identity for the action is
\begin{equation}
	\begin{split}\label{PolyakovWighman}
		S[\tilde g\tilde h] \ = \  S[\tilde g]+S[\tilde h]	+\frac{1}{\pi}\int_\Sigma d\tau d\sigma\ \str( \del hh^{-1}\ g^{-1}\bar\del g) \ .
\end{split}
\end{equation}


We define the chiral model action as
\begin{equation}
 S_{G\setminus H}[g]\ = \ S[g\omega(g^{-1})] \,.
\end{equation}
Clearly the action is invariant under right translation by $H$, i.e. 
\begin{equation}
 S_{G\setminus H}[gh] \ = \ S_{G\setminus H}[g]\ ,
\end{equation} where $h$ is a map from $\Sigma$ to $H$.
Using the fact that $\omega$ preserves the metric, as well as the Polyakov-Wiegmann identity and 
\begin{equation}
 S_{\text{WZ}}[\tilde{g}]+ S_{\text{WZ}}[\tilde{g}^{-1}]\ = \ 0 \qquad,\qquad S_{\text{kin}}[g]- S_{\text{kin}}[g^{-1}]\ = \ 0 
\end{equation}
we get
\begin{equation}
 S_{G\setminus H}\ = \ \frac{1}{\pi}\int d\tau d\sigma\  \str( g^{-1}\del g\ g^{-1}\bar{\del}g) - 
\frac{1}{\pi}\int d\tau d\sigma\  \str( \omega(g^{-1}\del g)\ g^{-1}\bar{\del}g)\, .
\end{equation}
We will now use this explicit form for the coset action to rewrite the Berkovits action of the deformed fermionic coset.
Note that a necessary condition for $G\setminus H$ to be conformal is that the Killing form (that is the supertrace in the adjoint)
of the underlying Lie superalgebra of $G$ is identically zero. This is the case for the algebra of the Berkovits fermionic coset, that is $\SLSA{sl}{4}{4}$.

\subsection{The Fermonic Coset}

We now rewrite the action of Berkovits (eq.(4.5) in \cite{Berkovits:2007zk}). There, world-sheet bosons were added to the fermionic coset and coupled non-trivially 
to the fermions.
The first term of the action of Berkovits is just the fermionic coset of the last section, the remaining terms are the world-sheet bosons and a term coupling bosons and fermions, i.e.
\begin{equation}
\mathcal L \ = \ \mathcal L_{\text{bos}} + \mathcal L_{\beta} + \mathcal L_{\text{int}}\, .
\end{equation}
We will now explicitly describe this Lagrangian.
Details of the Lie superalgebra $\SLSA{gl}{n}{n}$ and thus its subalgebra $\SLSA{sl}{n}{n}$ are given in the appendix. 
First recall the Gauss decomposition of $\SLSA{sl}{4}{4}$
\begin{equation}
\SLSA{sl}{4}{4} \ = \ \g_-\oplus\g_0\oplus \g_+,
\end{equation}
where $\g_0$ is the bosonic subgroup, and $\g_\pm$ are fermionic representations of $\g_0$. 
The automorphism describing the fermionic coset leaves $\g_0$ invariant and multiplies each fermion by minus one.
We now parameterize a group valued field $g=e^{\chi_+}e^{\chi_-}g_0$. 
That is, $\chi_\pm$ take values in the Grassmann envelope of $\g_\pm$ and $g_0$ parameterizes the bosonic subgroup.
Note that the invariant measure in this parameterization is
\begin{equation}
\mu(g) \ = \ d\chi_+d\chi_-\mu(g_0)
\end{equation}
and hence in the coset, i.e. after quotienting by $g_0$ the measure becomes the free one for the $\chi_\pm$.
We compute 
\begin{equation}
e^{-\chi_-}e^{-\chi_+}de^{\chi_+}e^{\chi_-} \ = \ d\chi_+ + [d\chi_+,\chi_-]+\frac{1}{2}[[d\chi_+,\chi_-],\chi_-]+d\chi_-\, .
\end{equation}
Hence
\begin{equation}
\begin{split}
\mathcal L_{\text{bos}} \ = \ &\str(\del\chi_+,\bar\del\chi_-) + \str( \del\chi_-,\bar\del\chi_+)+\\
&\frac{1}{2} \str( \del\chi_+, [[\bar\del\chi_+,\chi_-],\chi_-]) + 
\frac{1}{2} \str( \bar\del\chi_+, [[\del\chi_+,\chi_-],\chi_-])
\end{split}
\end{equation}
Define world-sheet bosons $\beta_\pm$ and $\bar\beta_\pm$ with Lagrangian
\begin{equation}
\mathcal L_{\beta} \ = \ \str(\beta_+\bar\del\beta_-)+\str(\bar\beta_+\del\bar\beta_-) +\str(\{\beta_+,\beta_-\},\{\bar\beta_+,\bar\beta_-\})\, .
\end{equation}
Finally the term coupling fermions to bosons is
\begin{equation}
\mathcal L_{\text{int}}\ = \ \str(\{\beta_+,\beta_-\},[\bar\del\chi_+,\chi_-]) + \str(\{\bar\beta_+,\bar\beta_-\},[\del\chi_+,\chi_-])\, .
\end{equation}
Writing the traces and brackets explicitly one recovers the Berkovits action.
The Lagrangian $\mathcal L$ splits into a free and an interacting part. 
The free one is 
\begin{equation}
\mathcal L_0 \ = \  \str( \del\chi_+,\bar\del\chi_-) + \str( \del\chi_-,\bar\del\chi_+)+
 \str(\beta_+\bar\del\beta_-)+\str(\bar\beta_+\del\bar\beta_-)\ ,
 \end{equation}
 and the interacting one is 
 \begin{equation}
\begin{split}
 \mathcal L_{J\bar  J} \ = \ &\frac{1}{2} \str(\del\chi_+, [[\bar\del\chi_+,\chi_-],\chi_-]) + 
\frac{1}{2} \str(\bar\del\chi_-, [[\del\chi_+,\chi_-],\chi_-]) +\\
&\str(\{\beta_+,\beta_-\},\{\bar\beta_+,\bar\beta_-\}) +\\
&\str(\{\beta_+,\beta_-\},[\bar\del\chi_+,\chi_-]) + \str(\{\bar\beta_+,\bar\beta_-\},[\del\chi_+,\chi_-])\, .
\end{split}
\end{equation}
The complete Lagrangian is
\begin{equation}
\mathcal L \ = \  \mathcal L_0 + \mathcal L_{J\bar  J}\,.
 \end{equation}
The action is
\begin{equation}
S\ = \ \frac{R^2}{\Lambda}\int d^2z\ \mathcal L
\end{equation}
hence after an appropriate rescaling it takes the form
\begin{equation}
S\ = \ \int d^2z\ \mathcal L_0 + \frac{\Lambda}{R^2}\mathcal L_{J\bar J}\, .
\end{equation}
We will now find symmetries of this model.

\section{Symmetries}

A natural way to study symmetries of an interacting theory that is a perturbation of a free theory is
to find those symmetries of the free model that are preserved by perturbation.
We thus start with the free theory, that is the large radius limit $R^2/\Lambda\rightarrow\infty$.
Here and in what follows, we sum over repeated indices unless otherwise stated.

\subsection{Large Radius Limit }

We summarize Lie superalgebra properties in Appendix \ref{app:glnn}.
The coset fields are fermionic fields $\chi^\pm$ in $\g_\pm$ and the
corresponding world-sheet bosons $\beta^\pm$. The components are defined as $\chi^\pm=\chi^\pm_{\al\be}F^{\al\be}_\pm$, where $\{F^{\al\be}_\pm\}$ is a basis of the fermionic part $\g_\pm$ of $\SLSA{sl}{4}{4}$. Their operator product expansions as fields in the free theory described by the Lagrangian $\mathcal L_0$ are
\begin{equation}
 \chi^+_{\al\be}(z)\chi^-_{\ga\de}(w) \ \sim \ \delta_{\al\de}\delta_{\be\ga}\ln|z-w|^2 
\end{equation}
 and 
\begin{equation}
 \beta^+_{\al\be}(z)\beta^-_{\ga\de}(w) \ \sim \ \frac{\delta_{\al\de}\delta_{\be\ga}}{(z-w)}\ \sim \ -\beta^-_{\al\be}(z)\beta^+_{\ga\de}(w)\, . 
\end{equation}
The anti-holomorphic quantites $\bar\beta^\pm$ are analogous. 
For us it is convenient to split $\chi_-$ in a locality-violating way
\begin{equation}
\chi_- \ = \ \chi_-^L+\chi_-^R\, .
\end{equation}
To simplify notation and to make contact to the standard $bc$-ghost systems, we define 
\begin{equation}
\chi_-^L\ = \ c\quad, \quad \del\chi^+\ = \ b \quad,\quad \chi_-^R \ = \ \bar c\quad,\quad \bar\del\chi_+\ = \ \bar b\, .
\end{equation}
The operator product of the components of the fields $b$ and $c$
are 
\begin{equation}
b_{\al\be}(z)c_{\ga\de}(w) \ \sim \ \frac{\delta_{\al\de}\delta_{\be\ga}}{(z-w)}\, . 
\end{equation}
We will restrict to the holomorphic currents and symmetries, as the anti-holomorphic part is in complete analogy.

\subsection{$\SLSA{pgl}{4}{4}$ currents at critical level}\label{sec:pglcurrents}

Using the free fields one can construct $\SLSA{pgl}{4}{4}$ currents of level zero. This construction is a mix
of Wakimoto and free field from fields in the adjoint representation of $\SLA{gl}{4}$. Formulae follow from \cite{Creutzig:2010ne}; see also Appendix \ref{app:glnn}. We choose the basis given in the appendix.
Introduce the currents
\begin{equation}\label{eq:sl4currents}
\begin{split}
J^{E_{+}^{\al\be}}_B \ &= \ \beta^+_{\be\ga}\beta^-_{\ga\al} \\
J^{E_{-}^{\al\be}}_B \ &= \ -\beta^-_{\be\ga}\beta^+_{\ga\al} \\
\end{split}
\end{equation}
These form two commuting copies of the level $-4$ current algebra of  $\SLA{gl}{4}$, where the central elements are related by
\begin{equation}\label{eq:relbos}
J^{E_{+}^{\al\al}}_B+J^{E_{-}^{\al\al}}_B\ = \ 0\, .
\end{equation} 
These currents can be used to get a $\SLSA{pgl}{4}{4}$ current algebra at level zero.
We introduce the currents
\begin{align}\label{eq:pglnncurrents}
    \begin{split}
      J^{E^{\al\be}_{\ep}} &= J^{E_{\ep}^{\al\be}}_B-\de_{\ep,+} b_{\be\ga}c_{\ga\al}+\de_{\ep,-}b_{\ga\al}c_{\be\ga}\, , \\
       J^{F_-^{\al\be}} &=-b_{\be\al} \, ,\\
       J^{F_{+}^{\al\be}} &=-c_{\be\ga}J^{E_{+}^{\al\ga}}_B-c_{\ga\al}J^{E_{-}^{\ga\be}}_B-b_{\ga\de}c_{\be\ga}c_{\de\al}.
    \end{split}
\end{align}
The computation of the operator product algebra is straightforward.
As a result, the above fields are $\SLSA{pgl}{4}{4}$ currents at level zero. 
The $\SLA{u}{1}$ current $J^{E^{\al\al}_{+}}-J^{E^{\al\al}_{-}}$ is a derivation, i.e. it does not appear on the right hand side of any current OPE. 
The derived subalgebra, that is the current subalgebra consisting of those currents that appear on the right hand side is $\SLSA{psl}{4}{4}$.
We remark, that we are working with the complex Lie superalgebra $\SLSA{psl}{4}{4}$. In Appendix \ref{app:real} we recall how to
obtain the real form $\SLSA{psu}{2,2}{4}$.

Interestingly the interaction term takes the simple current-current form
\begin{equation}
\mathcal L_{J\bar  J} \ = \ \text{str}(J,\bar J)
\end{equation}
where 
\begin{equation}
J\ = \ b+\{\beta^-,\beta^+\}+[c,b]+[\{\beta^-,\beta^+\},c]+\frac{1}{2}[c,[c,b]]
\end{equation}
and the components $J^{t_a}=\text{str}(t_a J)$, for $t_a$ in $\{E^{\al\be}_\pm,F^{\al\be}_\pm\}$, are exactly those of \eqref{eq:pglnncurrents}.
Note, that here it does not matter whether we include the $\SLA{u}{1}$ current as it is the isotropic element of $\SLSA{pgl}{4}{4}$.
Hence the perturbation is a $\SLSA{psl}{4}{4}$ current-current perturbation. 
Since the supertrace in the adjoint representation vanishes identically, the perturbed theory is one-loop conformal
invariant.

We would like to remark that the situation here is different from constructions of $\SLSA{psl}{4}{4}$ representations via $\SLSA{sl}{4}{4}$, where one has to restrict to special
representations to ensure that the ideal one is quotienting by acts trivially, and thus representations descend to the quotient.

\subsection{$\mathcal N=4$ superconformal algebra}

There is more symmetry in the free theory, namely the two supercurrents
\begin{equation}
G^+ \ = \ \beta^-_{\al\be}b_{\be\al}\qquad,\qquad G^- \ = \ \beta^+_{\al\be}c_{\be\al},
\end{equation}
which together with
\begin{equation}
\begin{split}
K_- \ = \ \beta^-_{\al\be}\beta^-_{\al\be} \ , \qquad 
K_3 \ = \ \beta^+_{\al\be}\beta^-_{\be\al} \ , \qquad
K_+ \ = \  \beta^+_{\al\be}\beta^+_{\al\be} 
\end{split},
\end{equation}
generate an $\mathcal N=4$ superconformal algebra. The $K_\pm, K_3$ generate an $\AKMA{sl}{2}$ of level $-8$.
Define 
\begin{equation}
G^+_0 \ = \ \frac{1}{2\pi i} \oint dz\, G^+(z) \, .
\end{equation}
Then one computes
\begin{equation}
\begin{split}
G^+_0 \beta^+_{\al\be}(w) \ &= \ J^{F^{\al\be}_-}(w) \qquad,\qquad G^+_0 \beta^+_{\ga\de}c_{\de\al}c_{\be\ga}(w) \ = \ -J^{F^{\al\be}_+}(w) \\
G^+_0 \beta^+_{\ga\al}c_{\be\ga}(w) \ &= \ -J^{E^{\al\be}_-}(w) \,\quad,\qquad \quad\,
G^+_0 \beta^+_{\be\ga}c_{\ga\al}(w) \ = \ J^{E^{\al\be}_+}(w) \\
\end{split}
\end{equation}
Hence, the current-current term is $G^+$ exact and thus preserves superconformal symmetry. 
In \cite{Berkovits:2007zk} the physical states were the cohomology for supercharges of this superconformal algebra.
But there is another interesting topological field theory.

\subsection{Critical Level Topological Algebra}

We now explain that the $\SLSA{gl}{4}{4}$ current algebra at non-critical level gives rise to a topological algebra.
This cannot be true anymore in the critical level limit as in this case there is no Virasoro field since
the Sugawara field becomes central. We thus get a modified topological algebra, i.e. the critical level limit
of topological algebras. 

In \cite{Creutzig:2010ne} a Sugawara-like construction of topological algebra from $\SLSA{gl}{n}{n}$ currents at level $k$
was given. 
The result is as follows; for notation see Appendix A.
\begin{equation}
 \begin{split}\label{eq:topalg}
    G_k^+\ &=\ \sum_{\alpha}J_k^{F^{\alpha\alpha}_{+}} \\
    U_k \ &=\ \frac{1}{2k}\sum_{\alpha}(k+n+1-2\alpha)J_k^{E_{+}^{\alpha\alpha}}-\frac{1}{2k}\sum_{\alpha}(k+2\alpha -n-1)J_k^{E_{-}^{\alpha\alpha}}\\
    G_k^-\ &=\ \frac{1}{2k}\sum_{\alpha}J_k^{F_{-}^{\alpha\alpha}}\Big(J_k^{E^{\alpha\alpha}_{+}}-J_k^{E^{\alpha\alpha}_-}+\frac{1}{k}\sum_\beta J_k^{E^{\beta\beta}_+}+J_k^{E^{\beta\beta}_-}\Big)-\frac{1}{k}\sum_{\alpha>\beta}J_k^{{F}_{-}^{\alpha\beta}}J_k^{E^{\beta\alpha}_-}\\
&\quad+\frac{1}{k}\sum_{\alpha<\beta}J_k^{F_-^{\alpha\beta}}J_k^{E^{\beta\alpha}_+}-\frac{1}{2k}\sum_\alpha(2N-2\alpha+1)\del J_k^{F_-^{\alpha\alpha}}\, .\\
\end{split}
\end{equation}
Note that these expressions are normally ordered.
The operator product algebra of these currents is that of a $c=0$ twisted superconformal, i.e. topological, algebra:
\begin{equation}
\begin{split}\label{eq:defitopope}
G_k^+(z)\, G_k^-(w)& \sim \frac{U_k(w)}{(z-w)^2} + \frac{T_k(w)}{(z-w)}\, ,
 \\[2mm]
U_k(z) \, G_k^\pm(w) & \sim   \frac{\pm G_k^\pm(w)}{(z-w)}\quad ,\ \
U_k(z)\, U_k(w) \, \sim \,  0\, .
\end{split}
\end{equation}
Here $T_k$ is the Virasoro field of $\SLSA{gl}{n}{n}$.
From the explicit expressions \eqref{eq:topalg}, we see that the limit $k\rightarrow 0$ is not well-defined.
We thus need to rescale, i.e. we define
\begin{equation}
G^+ \ = \ \lim_{k\rightarrow 0} G^+_k \quad,\quad
G^- \ = \ \lim_{k\rightarrow 0} kG^-_k \quad,\quad
U \ = \ \lim_{k\rightarrow 0} kU \quad,\quad 
S \ = \ \lim_{k\rightarrow 0} kT_k \, .
\end{equation}
and we set the central $\SLA{u}{1}$ to zero
\begin{equation}\label{eq:limitzero}
0 \ = \ \lim_{k\rightarrow 0}\sum_\beta J_k^{E^{\beta\beta}_+}+J_k^{E^{\beta\beta}_-}
\end{equation}
The $S$ is for Sugawara-field.
Equation \eqref{eq:limitzero} is to ensure that we descend to $\SLSA{pgl}{n}{n}$, since in the critical level limit we can set the central $\SLA{u}{1}$ current of $\SLSA{gl}{n}{n}$ to zero as it has vanishing operator product expansion with any other field in this limit.

Note that these rescaled quantities are finite in the limit, and can now straightforwardly be expressed in
terms of our level zero $\SLSA{pgl}{4}{4}$ currents in the case $n=4$.
These fields now satisfy a critical level limit of a topological algebra, that is
\begin{equation}
\begin{split}\label{eq:crittop}
G^+(z)\, G^-(w)& \sim \frac{U(w)}{(z-w)^2} + \frac{S(w)}{(z-w)}\quad , \quad 
U(z)\, U(w) \, \sim \,  U(z) \, G^\pm(w) \, \sim \, 0\, .
\end{split}
\end{equation}
Note that as usual in a critical level limit, the Sugawara-field becomes central. We conclude this subsection with a remark on the cohomology.

The complete analogous construction of topological algebra holds for the anti-chiral part. Denote the corresponding fields with a bar.
The zero mode $G^+_0+\bar G^+_0$ then defines the critical level limit of a BRST-charge. It further is a good charge for 
the fermionic coset as the
perturbation $\mathcal L_{J\bar J}$ is in its kernel and thus it also makes sense to study its cohomology in the perturbed theory. 

We compute that $b_{\al\al},\beta^\pm_{\al\al}$ and $c_{\al\al}$ represent non-trivial cohomology classes.
Recall that $b=\del\chi^+$ and certainly also
$\chi^\pm_{\al\al}$ are BRST-closed but not BRST-exact.
Thus, the physical spectrum of this modified topological algebra contains
the $\SLSA{psl}{1}{1}$ WZW model, usually called symplectic fermions, with world-sheet supersymmetry.

Similarly one can show that the generators $C_1, D_\pm$ of the $\mathcal{W}$-algebra we are going to introduce in the next section are non-trivial in cohomology.

\subsection{Critical Level $\mathcal{W}^{(2)}_4$-Algebra}

So far, we have seen that the coset is a current-current perturbation of $\SLSA{psl}{4}{4}$ currents at critical level. 
Certainly OPEs of fields that commute with the perturbation remain unchanged. Further critical level affine algebras admit large centers, so we expect to find an interesting algebra preserved by the perturbation.

The problem to solve is to compute the commutant of $\SLSA{psl}{4}{4}$ at critical level inside the free theory.
Computing this commutant exhaustively requires hard work \cite{CGL}.
The first step is to show that the commutant equals the commutant of $\AKMA{sl}{4}_{-4}\oplus \AKMA{sl}{4}_{-4}$ (see \eqref{eq:sl4currents}) in the $\beta_\pm$ theory. Note, that $-4$ is the critical
level of $\AKMA{sl}{4}$.
This commutant is then generated by $C_1, C_2,C_3, D_\pm$ \cite{CGL}. 
The $C_i$ are the traces of the matrices $(\beta_+\beta_-)^i$ and the $D_\pm$ are the determinants of the $\beta_\pm$. The only non-trivial OPE relations are:
\begin{equation}\label{eq:wope}
\begin{split}
C_1(z) C_1(w) \ &\sim\ - \frac{16}{(z-w)^2}\qquad,\qquad
C_1(z) D_\pm(w)\ \sim\ \pm  \frac{4 D_\pm(w)}{(z-w)}\\
D_-(z) D_+(w)\ &\sim\ -\frac{24}{(z-w)^4}  - \frac{6 C_1(w)}{ (z-w)^3}  + \frac{A_2(w)}{(z-w)^2}+ \frac{A_1(w)}{(z-w)},
\end{split}
\end{equation} 
where 
\begin{equation}
\begin{split}
A_1(w)\ &= \  -\frac{1}{8} C_3(w) -\frac{1}{4} :C_2(w) C_1(w): + \frac{1}{16} :C_1(w) C_1(w) C_1(w): + \\ 
&\quad+ \frac{3}{4} :C_1(w)
\partial C_1(w): + \partial^2 C_1(w) \\
A_2(w)\ &= \ -C_2(w) + \frac{3}{4} :C_1(w) C_1(w): +3 \partial C_1(w).
\end{split}
\end{equation}
These are the OPEs of the $\mathcal{W}^{(2)}_4$-algebra at critical level $-4$. This $\mathcal{W}$-algebra can be conjecturally constructed from $\AKMA{sl}{4}$ via Drinfeld-Sokolov reduction. For this one has to start with an embedding of $\SLA{sl}{2}$ in $\SLA{sl}{4}$ such that as an $\SLA{sl}{2}$-representation, $\SLA{sl}{4}$ decomposes into 
\begin{equation}
\SLA{sl}{4} \ = \ \underline{1}\oplus\underline{3}\oplus \underline{3}\oplus\underline{3}\oplus\underline{5}\, .
\end{equation} 
At critical level the $\mathcal{W}^{(2)}_4$-algebra is generated by $C_1, C_2,C_3, D_-, D_+$ with the above OPEs plus an additional field $C_4$ that commutes with all others. 
In our case we have a non-trivial relation among these generators
\begin{equation}
\begin{split}
0\ &= \ :D_- D_+: - \frac{1}{256} C_4 + \frac{1}{32} :C_1 C_3: + \frac{1}{32} :C_2 C_1  C_1: - 
 \frac{1}{256} :C_1 C_1 C_1 C_1: + \\ &\quad\quad+\frac{1}{8} :\partial C_1 C_2: 
  -\frac{3}{32} : \partial C_1 C_1 C_1: - 
 \frac{1}{4} :\partial^2 C_1 C_1: - \frac{3}{16} :\partial C_1 \partial C_1: - \frac{1}{4} \partial^3 C_1 \, ,
\end{split}
 \end{equation} which allows us to eliminate $C_4$. Moreover, it was shown in \cite{CGL} that there are no non-trivial relations among the remaining fields $C_1, C_2,C_3, D_-, D_+$ and their derivatives, so they freely generate our commutant.
 
The Zhu map is a functor from the category of chiral conformal field theories to associative algebras. 
In \cite{CGL}, we related the commutant to a
classical ring of invariant differential operators using the Zhu functor.
We will now use the Zhu functor to map the theories and symmetries 
we found to higher spin algebras.

\section{A Higher Spin Algebra for the Free Limit}

Our main motivation was to find a relation between higher spin symmetries and the world-sheet sigma model in the free limit.
In this section, we will first introduce the higher spin algebra. This is a Weyl or oscillator algebra. 
We will then see that this type of higher spin algebra is related to the large radius limit of the 
sigma model via the Zhu map.
Finally, we are going to study the action of $\SLSA{psl}{4}{4}$ 
on this algebra.

\subsection{The Oscillator Algebra}

Define $16$ bosonic coordinates $x_{\al\be}$ and $16$ fermionic coordinates $\theta_{\al\be}$
and their derivatives $x_{\al\be}^\dag=\frac{d}{dx_{\be\al}}$ and $\theta_{\al\be}^\dag=\frac{d}{d\theta_{\be\al}}$ where $\al,\be=1,2,3,4$.
We have the commutation relations 
\begin{equation}
 [x_{\al\be}^\dag, x_{\ga\de}]\ = \ \delta_{\be\ga}\delta_{\al\de}\quad\text{and} \quad 
\{\theta_{\al\be}^\dag, \theta_{\ga\de}\}\ = \ \delta_{\be\ga}\delta_{\al\de}\,.
\end{equation}
Restricting to the bosonic coordinates, we can define 
\begin{equation}
 K^+_{\al\be} \ = \ x_{\be\ga}x^\dag_{\ga\al}\quad\text{and}\quad K^-_{\al\be} \ = \ -x_{\ga\be}x^\dag_{\al\ga}\, .
\end{equation}
These generate two
commuting copies of $U(\SLA{gl}{4})$ as associative subalgebras of the Weyl
algebra. The commutant of these two algebras inside our Weyl algebra is the associative algebra generated by the
$\SLA{u}{1}$, $c_1=-K^+_{\al\al}=K^-_{\al\al}$, the Casimirs
$c_2=K^-_{\al\be}K^-_{\be\al}$, $c_3=K^-_{\al\be}K^-_{\be\ga}K^-_{\ga\al}$, $c_4=K^-_{\al\be}K^-_{\be\ga}K^-_{\ga\de}K^-_{\de\al}$
and the two determinants $d^\dag=$ det$(x^\dag)$, $d=$ det$(x)$.
Their commutation relations are:
\begin{equation}
\begin{split}
[c_1,d]\ &= \ 4d\qquad,\qquad
[c_1,d^\dag]\ = \ -4d^\dag\\
[d,d^\dag]\ &= \ -24-6c_1 -c_2 + \frac{3}{4} c_1c_1 
 -\frac{1}{8} c_3-\frac{1}{4}c_2c_1 + \frac{1}{16}c_1c_1c_1 
\end{split}
\end{equation}
Comparing with (\ref{eq:wope}) we see this is the higher spin analogue of our critical level $\mathcal{W}^{(2)}_4$-algebra. As shown in \cite{CGL}, this algebra coincides with the Zhu algebra of $\mathcal{W}^{(2)}_4$, and we denote it by ${\cal Z}^{(2)}_4$.
As with the critical level $\mathcal{W}^{(2)}_4$-algebra, this algebra is not freely generated, since we have non-trivial relations:
\begin{equation}
\begin{split}
0 \ &= \ dd^\dag -\frac{1}{256} c_4 + \frac{1}{32}  c_1 c_3 + \frac{1}{32}  c_2 c_1^2 - \frac{1}{256}  c_1^4 + 
 \frac{1}{2} c_3 + \frac{7}{8} c_2 c_1 - \frac{5}{32} c_1^3 + 6 c_2 + \\
 &\qquad - \frac{35}{16} c_1^2 - \frac{25}{2} c_1 - 24\,,\\
0\ &= \ d^\dag d -\frac{1}{256} c_4 + \frac{1}{32}  c_1 c_3 + \frac{1}{32}  c_2 c_1^2 - \frac{1}{256}  c_1^4 + \frac{3}{8} c_3 + \frac{5}{8} c_1 c_2 - \frac{3}{32}c_1^3 + 3 c_2 +\\
&\qquad - \frac{11}{16} c_1^2 -  \frac{3}{2}\,.
\end{split}
\end{equation}
We can also easily realize the algebra $\SLSA{pgl}{4}{4}$ with the oscillators 
\begin{align}\label{eq:pglalgebra}
    \begin{split}
      E^{\al\be}_{\ep} &= K^{\ep}_{\al\be}-\de_{\ep,+} \theta^\dag_{\be\ga}\theta^{\ga\al}+\de_{\ep,-}\theta^\dag_{\ga\al}\theta^{\be\ga}\, , \\
       F_-^{\al\be} &=-\theta^\dag_{\be\al} \, ,\\
       F_{+}^{\al\be} &=-\theta^{\be\ga}K^{+}_{\al\ga}-\theta^{\ga\al}K^{-}_{\ga\be}-\theta^\dag_{\ga\de}\theta^{\be\ga}\theta^{\de\al}.
    \end{split}
\end{align}
where $\epsilon=\pm$. 

\subsection{The Zhu map}

We relegate the precise definition of the Zhu functor to Appendix \ref{app:zhu}. The Zhu functor is a functor from the category of
chiral conformal field theories to the category of associative algebras.
First, we relate the free theory, that is the large radius limit of the world-sheet sigma model, to its higher-spin/Zhu algebra.
Here, we should note that we always restrict to fields expressed through $\beta^\pm,c=\chi^-_L$ and $b=\del\chi^+$, i.e. 
we neglect $\chi^+$. 
The Zhu map is straightforward for the free conformal field theory
\begin{equation}\label{eq:zhufree}
\begin{split}
\pi_{\text{zhu}}(\beta^+_{\al\be})\ &= \ x^\dag_{\al\be}\ \ , \ \ 
\pi_{\text{zhu}}(\beta^-_{\al\be})\ = \ x_{\al\be}\ \ , \ \ 
\pi_{\text{zhu}}(b_{\al\be})\ = \ \theta^\dag_{\al\be}\ \ , \ \ 
\pi_{\text{zhu}}(c_{\al\be})\ = \ \theta_{\al\be}\, .
\end{split}
\end{equation}
This basic Zhu map allows to compute all the other related maps. For example, we have the affine Lie algebra $\AKMA{gl}{4}_{-4}\oplus\AKMA{gl}{4}_{-4}/\AKMA{gl}{1}$, where the $\AKMA{gl}{1}$ denotes the quotient by the relation \eqref{eq:relbos}. 
From (\ref{eq:zhufree}) and the expressions in (\ref{eq:sl4currents}) we compute the Zhu map for the affine Lie algebra 
\begin{equation}
\begin{split}
\pi_{\text{zhu}}: \AKMA{gl}{4}_{-4}\oplus\AKMA{gl}{4}_{-4}/\AKMA{gl}{1} &\rightarrow \SLA{gl}{4}\oplus\SLA{gl}{4}/\SLA{gl}{1} \quad,\quad 
\pi_{\text{zhu}}(J^{E_{\ep}^{\al\be}}_B) \ = \  K_{\al\be}^\epsilon\, .
\end{split}
\end{equation}
The map extends to the currents  of the complete superalgebra $\AKMSA{pgl}{4}{4}_{0}$ (\ref{eq:pglnncurrents}) as follows
\begin{equation}
\begin{split}
\pi_{\text{zhu}}: \AKMSA{pgl}{4}{4}_{0} &\rightarrow \SLSA{pgl}{4}{4} \quad,\quad
\pi_{\text{zhu}}(J^{E_{\ep}^{\al\be}}) \ = \  E^{\al\be}_\epsilon \quad,\quad
\pi_{\text{zhu}}(J^{F_{\ep}^{\al\be}}) \ = \ F^{\al\be}_\epsilon\ .
\end{split}
\end{equation}
The map for the $\mathcal{W}$-algebra is harder to compute, but only for the Casimirs.
We denote the degree of the Casimir element $c_i$ by $i$, and the degree of a product 
of Casimirs is the sum of the degrees of the factors. We get
\begin{equation}
\begin{split}
\pi_{\text{zhu}}: \mathcal{W}^{(2)}_4 &\rightarrow {\cal Z}^{(2)}_4 \\
\pi_{\text{zhu}}(C_1)\ = \ c_1 \ \ , \ \ 
\pi_{\text{zhu}}(D_+)\ = \ d\ \ , \ \ 
\pi_{\text{zhu}}(D_-) \ &= \ d^\dag \ \ , \ \ 
\pi_{\text{zhu}}(C_i) \ = \ c_i +\dots
\end{split}
\end{equation}
for $i=2,3,4$.
 The dots denote terms consisting of Casimirs of lower
degree, though $c_1$ never appears. Interested readers are recommended to consult \cite{CGL} for explicit expressions of the Casimirs.

\subsection{The Interaction term and a higher spin extension} 

So far, we have completely neglected the anti-chiral part of the world-sheet theory. The reason is that we can treat it in complete analogy.
For example, the anti-chiral oscillator-algebra is generated by $\bar x_{\al\be},\bar x^\dag_{\al\be},\bar \theta_{\al\be},\bar \theta^\dag_{\al\be}$
with relations
\begin{equation}
 [\bar x_{\al\be}^\dag, \bar x_{\ga\de}]\ = \ \delta_{\be\ga}\delta_{\al\de}\quad\text{and} \quad 
\{\bar \theta_{\al\be}^\dag, \bar \theta_{\ga\de}\}\ = \ \delta_{\be\ga}\delta_{\al\de}\,.
\end{equation}
Similarly, we have a second copy of the Lie algebra $\SLA{gl}{4}\oplus\SLA{gl}{4}/\SLA{gl}{1}$, where we
call the generators by $\bar K_{\al\be}^\epsilon$. We have anti-chiral
generators $\bar E^{\al\be}_\epsilon, \bar F^{\al\be}_\epsilon$ for the superalgebra $\SLSA{pgl}{4}{4}$ and the anti-chiral set of associative algebra generators, which we denote by
$\bar c_i, \bar d, \bar d^\dag$. 

Now, we can apply the Zhu map to the interaction term of the fermionic coset model, we call the image in the higher spin algebra $H$
\begin{equation}
H\ = \ \pi_{\text{zhu}}(\str(J,\bar J)) \ = \ 
E^{\al\be}_+\bar E^{\be\al}_+- E^{\al\be}_-\bar E^{\be\al}_- +
F^{\al\be}_-\bar F^{\be\al}_+- F^{\al\be}_+\bar F^{\be\al}_-\, .
\end{equation}
The Zhu map image of the algebra ${\cal Z}^{(2)}_4\oplus {\cal Z}^{(2)}_4$ has the crucial property that it commutes with the interaction operator $H$
\begin{equation}
[H,x]\ = \ 0 \qquad\text{for all} \ x\ \text{in} \ {\cal Z}^{(2)}_4\oplus {\cal Z}^{(2)}_4\, .
\end{equation}
But also the diagonal $\SLSA{psl}{4}{4}$ algebra generated by
\begin{equation}\label{eq:diag}
\mathcal{E}^{\al\be}_\pm\ = \ E^{\al\be}_\pm+\bar E^{\al\be}_\pm\qquad,\qquad \mathcal{F}^{\al\be}_\pm\ = \ F^{\al\be}_\pm+\bar F^{\al\be}_\pm
\end{equation}
commutes with $H$.
This means that our interaction preserves $\SLSA{psl}{4}{4}\oplus {\cal Z}^{(2)}_4\oplus {\cal Z}^{(2)}_4$-symmetry.
Note that the operators $d,d'$ have spin two, while the Casimirs $c_i$ have spin $i$. 
We will now study the representations of this extension of $\SLSA{psl}{4}{4}$.
Note that the finite-dimensional, irreducible representations of ${\cal Z}^{(2)}_4$ have been classified in \cite{CGL}.

\subsection{Representations of $\SLSA{psl}{4}{4}\oplus {\cal Z}^{(2)}_4\oplus {\cal Z}^{(2)}_4$}

In this section, by a $\SLSA{psl}{4}{4}$ action, we mean the one generated by the elements of \eqref{eq:diag}.
The algebra $\SLSA{psl}{4}{4}\oplus {\cal Z}^{(2)}_4 \oplus {\cal Z}^{(2)}_4$ naturally acts on the space of functions in the $32$ even
variables $x_{\al\be},\bar x_{\al\be}$ and the $32$ odd variables $\theta_{\al\be},\bar \theta_{\al\be}$.
Since we wish to have finite dimensional representations of $\SLSA{psl}{4}{4}$, we restrict to the polynomial ring $R[x,\bar x,\theta,\bar\theta]$ in these variables.
Recall the induced module construction for the type I Lie superalgebra $\SLSA{psl}{4}{4}$.
Let $M$ be a module of $\SLA{sl}{4}\oplus\SLA{sl}{4}$. Then letting the $F_+$ act trivially and the $F_-$ freely, one obtains a module of $\SLSA{psl}{4}{4}$, denoted by $\text{Ind}^{\SLSA{psl}{4}{4}}_{\SLA{sl}{4}\oplus \SLA{sl}{4}\oplus\g_+}\bigl(M\bigr)$.
We are going to show that
\begin{equation}
R[x,\bar x,\theta,\bar\theta] \ = \ \bigoplus_{m,n}\text{Ind}^{\SLSA{psl}{4}{4}}_{\SLA{sl}{4}\oplus \SLA{sl}{4}\oplus\g_+}\bigl(\text{Sym}^n(V_{(1,0,0;1,0,0)})\otimes \text{Sym}^m(V_{(1,0,0;1,0,0)})\bigr)
\end{equation}
Here $V_\Lambda$ denotes the highest-weight representation with highest-weight $\Lambda$, and conventions on
weights are given in Appendix \ref{app:weights}.
Further, we will find in the decomposition of the sum indecomposable but reducible representations.
The quotient of the maximal proper subrepresentations of these indecomposable ones leads to a direct sum
of so-called atypical representations. In physics these are often called short-multiplets or BPS-representations, as they possess a highest-weight state that in addition is also annihilated by some of the fermionic creation operators. In our case, we have at least $1/4$-BPS states.

Let us explain these statements. 
We start with representations of $\SLA{sl}{4}\oplus \SLA{sl}{4}$. Define the bi-degree of a monomial 
\begin{equation}
f \ = \ \prod x_{\al\be}^{n_{\al\be}}\bar x_{\ga\de}^{m_{\ga\de}}
\end{equation}
to be deg$(f)=(n,m)=(\sum n_{\al,\be},\sum m_{\ga\de})$, where the sums are understood to run over all even variables $x_{\al\be},\bar x_{\al\be}$.
Clearly the degree is invariant under the action of $\SLA{sl}{4}\oplus \SLA{sl}{4}$.
We call the module consisting of all functions of degree $(n,m)$ the $M^a_{n,m}$.
Functions of degree $(1,0)$ transform in the representation $V_{(1,0,0;1,0,0)}$, where
$(1,0,0)$ is the highest-weight of the standard representation of $\SLA{sl}{4}$.
More generally, we have
\begin{equation}
M^a_{n,m}\ =\ \text{Sym}^n(V_{(1,0,0;1,0,0)})\otimes \text{Sym}^m(V_{(1,0,0;1,0,0)})\, .
\end{equation}
Next, we want to add the fermionic coordinates. We consider the element
\begin{equation}
\lambda \ = \ \prod (\theta_{\al\be}+\bar \theta_{\al\be})\, .
\end{equation}
where the product runs over all fermionic variables. 
It transforms in the trivial representation of $\SLA{sl}{4}\oplus \SLA{sl}{4}$.
We define the $\SLA{sl}{4}\oplus \SLA{sl}{4}$ module $M^b_{n,m}$ as the module generated by functions of degree $(n,m)$ times $\lambda$.
As an $\SLA{sl}{4}\oplus \SLA{sl}{4}$ module 
\begin{equation}
M^b_{n,m}\ =\ \text{Sym}^n(V_{(1,0,0;1,0,0)})\otimes \text{Sym}^m(V_{(1,0,0;1,0,0)})\, .
\end{equation}
Clearly, the $\mathcal F^{\al\be}_+$ action annihilates any function in $M^b_{n,m}$.
Recall, that the fermionic subalgebra generated by these $\mathcal F^{\al\be}_+$ elements we called $\g_+$.
We thus have promoted the module $M^b_{n,m}$ to be an $\SLA{sl}{4}\oplus \SLA{sl}{4}\oplus\g_+$-module.
Now, the fermionic elements $\mathcal F^{\al\be}_-$ act freely on $\lambda$ and hence on $M^b_{n,m}$. Thus as a $\SLSA{psl}{4}{4}$-module we have just the induced Kac module.
In summary, we have shown, that functions of degree $(n,m)$ and any degree in the odd variables form a $\SLSA{psl}{4}{4}$-module $M_{n,m}$ that is
isomorphic to the Kac module: 
\begin{equation}
M_{n,m} \ \cong \  \text{Ind}^{\SLSA{psl}{4}{4}}_{\SLA{sl}{4}\oplus \SLA{sl}{4}\oplus\g_+}\bigl(\text{Sym}^n(V_{(1,0,0;1,0,0)})\otimes  \text{Sym}^m(V_{(1,0,0;1,0,0)})\bigr) \, .
\end{equation}
These Kac modules are finite-dimensional and in general reducible.
We will later see, that we are in non-generic position, so our Kac modules are reducible but indecomposable.

Now, let us state how the algebra extension to $\SLSA{psl}{4}{4}\oplus {\cal Z}^{(2)}_4\oplus {\cal Z}^{(2)}_4$ combines these modules.
The non-trivial generators of ${\cal Z}^{(2)}_4\oplus {\cal Z}^{(2)}_4$ define maps
\begin{equation}
\begin{split}
d:M_{n,m} \ &\longrightarrow \ M_{n+4,m} \qquad,\qquad d^\dag: M_{n,m} \ \longrightarrow \ M_{n-4,m} \\
\bar d:M_{n,m} \ &\longrightarrow \ M_{n,m+4} \qquad,\qquad \bar d^\dag: M_{n,m} \ \longrightarrow \ M_{n,m-4} \\
\end{split}
\end{equation}
The maps $d,\bar d$ are injective and not surjective, while $d^\dag,\bar d^\dag$ are surjective and not injective.

It remains to show that some of the induced modules we found are non-generic.
For this we will decompose $\text{Sym}^n(V_{(1,0,0;1,0,0)})$ into irreducible representations. 
Therefore, we need to compute highest-weight vectors. It turns out that all these vectors can
be built from four polynomials. Let $i\in\{1,2,3,4\}$, define $i\times i$ matrices $B^i$ whose entries are 
$B^i_{ab}=x^{5-a,b}$.
Then $v_i=\text{det} B^i$ is a highest-weight vector and the corresponding weights are
\begin{equation}
\begin{split}
\Lambda_1\ &= \ (1,0,0;1,0,0)\qquad , \qquad 
\Lambda_2\ = \ (0,1,0;0,1,0)\, , \\
\Lambda_3\ &= \ (0,0,1;0,0,1)\qquad , \qquad
\Lambda_4\ = \ (0,0,0;0,0,0)\, .
\end{split}
\end{equation} 
Let $a_1+2a_2+3a_3+4a_4=n$, then clearly $v_1^{a_1}v_2^{a_2}v_3^{a_3}v_4^{a_4}$ 
is a highest-weight vector of $\text{Sym}^n(V_{(1,0,0;1,0,0)})$.
We claim that these are all. This can be proven by comparing dimensions, which we did up to $n=6$.
In summary
\begin{equation}
\text{Sym}^n(V_{(1,0,0;1,0,0)}) \ = \ \bigoplus_{\sum ia_i=n} V_{(a_1,a_2,a_3;a_1,a_2,a_3)}\, .
\end{equation}
We list the first examples
\begin{equation*}
\begin{split}
\text{Sym}^0(V_{(1,0,0;1,0,0)})\ &= \ V_{(0,0,0;0,0,0)} \\
\text{Sym}^1(V_{(1,0,0;1,0,0)})\ &= \ V_{(1,0,0;1,0,0)} \\
\text{Sym}^2(V_{(1,0,0;1,0,0)})\ &= \ V_{(2,0,0;2,0,0)} \oplus V_{(0,1,0;0,1,0)} \\
\text{Sym}^3(V_{(1,0,0;1,0,0)})\ &= \ V_{(3,0,0;3,0,0)} \oplus V_{(1,1,0;1,1,0)} \oplus V_{(0,0,1;0,0,1)} \\
\text{Sym}^4(V_{(1,0,0;1,0,0)})\ &= \ V_{(4,0,0;4,0,0)} \oplus V_{(2,1,0;2,1,0)} \oplus V_{(1,0,1;1,0,1)}\oplus V_{(0,0,0;0,0,0)} \oplus V_{(0,2,0;0,2,0)} \\
\end{split}
\end{equation*}
Recall, that the Kac module is irreducible whenever the inner product of the highest-weight shifted by the Weyl vector with any odd isotropic
root is non-zero. This is the generic case and thus called typical representation.
Otherwise the Kac module contains proper submodules. The quotient by a maximal one is then the atypical irreducible representation.
Looking at the appendix \eqref{eq:atyp} we see that we indeed are in non-generic position and taking corresponding quotients, we obtain atypical short-multiplets.

\section{Conclusion and Outlook}

In this paper, we have shown that the fermionic coset model introduced in \cite{Berkovits:2007zk} has a higher spin symmetry algebra $\mathcal{W}^{(2)}_4$ at critical level, preserved by world-sheet interaction. Moreover, in the large volume limit the linear A-model of  \cite{Berkovits:2007zk} contains a $\AKMSA{pgl}{4}{4}_0$ current algebra. Using the Zhu map, we can interpret these symmetry algebras as the world-sheet origin of higher spin algebras in AdS$_5\times$S$^5$ spacetime, namely a higher spin algebra extending $\SLSA{psl}{4}{4}$ and a subalgebra of it. While the former is the correct symmetry algebra for the free limit of large $N$ four dimensional ${\cal N}=4$ supersymmetric Yang-Mills theory, the latter is much larger than $\SLSA{psl}{4}{4}$ that remains when interaction is turned on in the Yang-Mills theory. 

One possible interpretation of our findings is that the difference in BRST operators between the fermionic coset and the pure spinor model is essential, as we have seen that they lead to different spectrum of operators in the bulk of the world-sheet. Taking this interpretation a bit further, we may infer that the right world-sheet topological twisting must have the same BRST charge as in the pure spinor model. This was indeed suggested in the later papers by Berkovits\cite{Berkovits:2008qc,Berkovits:2008ga}. In addition, it should be nonetheless interesting to consider boundary conditions in the fermionic coset model and work out the open sector spectrum for BRST cohomology. With the assumption that vertex operators can be constructed without major difficulty, it will unambiguously determine the spacetime theory for the open string sector which should be different from ${\cal N}=4$ supersymmetric Yang-Mills theory. It would be illuminating if a relation can be found between the two, especially interesting is the possibility that the former is a sub-sector of the latter (Yang-Mills) theory.


The observation that the nonlinear A-model is a perturbation of current-current type is surprising and remarkable. If it generalizes, it opens a further direction of study of sigma models on supermanifolds, as it directly relates the model to a theory with affine Lie super algebra as symmetry.
Conformal field theories of this type are the Wess-Zumino-Witten models of Lie super groups. 
These theories are fairly well-understood, due to their additional symmetry, see e.g. \cite{Rozansky:1992rx,Schomerus:2005bf,Saleur:2006tf,Gotz:2006qp,Quella:2007hr,Creutzig:2007jy,Creutzig:2009fh,Hikida:2007sz,Creutzig:2009zz,Creutzig:2008an,Creutzig:2008ag,Creutzig:2008ek,Creutzig:2010zp,Creutzig:2011cu,Creutzig:2011qm}.
Also perturbations of current-current type are explored. It is possible to compute the spectra
of perturbed theories \cite{Quella:2007sg,Candu:2009ep,Mitev:2008yt}, to compute some correlation functions \cite{Konechny:2010nq,Ashok:2009xx,Benichou:2010rk} and 
to establish Yangian symmetry \cite{Creutzig:2010hr}.
The nonlinear A-model is roughly speaking a world-sheet supersymmetric version of
the fermionic coset. 
The open and interesting question is, whether one can find other world-sheet supersymmetric super cosets, such that they can be described as perturbations of current-current type. 
Looking back, at the constructon of the action in section 2 this is certainly possible for cosets described by an involutive automorphism possessing a complex structure. More general super cosets should be studied.

\noindent\textbf{Acknowledgements.}
We would like to thank Yuri Aisaka, Luca Mazzucato, Thomas Quella, Balt van Rees and Volker Schomerus for useful conversations. 
The work of PG is supported in part by DOE grant DE-FG02-92ER-40697. 

\appendix

\section{Properties of $\SLSA{gl}{n}{n}$}\label{app:glnn}

In this section, we introduce properties of the Lie superalgebra $\SLSA{gl}{n}{n}$.
Notation and computations are mostly as in \cite{Creutzig:2010ne}.

\noindent A convenient basis for $\SLSA{gl}{n}{n}$ is  $\{ E_\ep^{\al\be}, F_\ep^{\al\be}\ | \ 1\leq
\al,\be \leq n,\, \ep=\pm\,\}$, where the generators $E$ are bosonic
 and $F$ are fermionic. The advantage of this notation is
that the invariant bilinear form and the super commutation relations are easy to express.
The metric takes the form
\begin{align}\label{eq:ksupertraceglnn}
    k\str(E^{\al\be}_{\ep}E^{\al'\be'}_{\ep'})&:=\ka^{\binom{\al\be}{\ep}\binom{\al'\be'}{\ep'}}=k\ep\de_{\ep\ep'}\de^{\al\be'}\de^{\be\al'}\, ,\\
    k\str(F^{\al\be}_{\ep}F^{\al'\be'}_{\ep'})&=k\varepsilon_{\ep\ep'}\de^{\al\be'}\de^{\be\al'},
\end{align}
where $\varepsilon_{\ep\ep'}$ is the antisymmetric symbol with
$\ep_{+-}=1$.

The non-vanishing Lie super algebra relations are
\begin{equation}\label{eq:glnngenerators}
 \begin{split}
  [E_\ep^{\al\be},E_{\ep'}^{\ga\de}] \ &= \ \de_{\ep,\ep'}\left(\delta^{\be\ga}E_\ep^{\al\de}- \delta^{\al\de}E_\ep^{\ga\be}\right)\, , \\
  [E_\ep^{\al\be},F_{\ep'}^{\ga\de}] \ &= \ \de_{\ep,\ep'}\delta^{\be\ga}F_\ep^{\al\de}-\de_{-\ep,\ep'}\delta^{\al\de}F_{\ep'}^{\ga\be}\, , \\
  \{F_\ep^{\al\be},F_{\ep'}^{\ga\de}\} \ &= \ \de_{-\ep,\ep'}\left(\delta^{\be\ga}E_\ep^{\al\de}+\delta^{\al\de}E_{\ep'}^{\ga\be}\right)\, . \\
 \end{split}
\end{equation}
We define the fermionic fields
\begin{align}\label{eq:bcmatrices}
    c&=c_{\al\be}F_-^{\al\be}, \\
    b&=b_{\al\be}F_+^{\al\be}.
\end{align}
These
satisfy the OPEs
\begin{equation}
c_{\be\al}(z)b_{\ga\de}(w) \ \sim \ \frac{\delta_{\al\ga}\delta_{\be\de}}{(z-w)}\, .
\end{equation}
The $\SLSA{gl}{n}{n}$ currents are then given by
\begin{align}\label{eq:glnncurrents}
    \begin{split}
      J^{E^{\al\be}_{\ep}} &= J^{E_{\ep}^{\al\be}}_B-\de_{\ep,+} b_{\be\ga}c_{\ga\al}+\de_{\ep,-}b_{\ga\al}c_{\be\ga}\, , \\
       J^{F_-^{\al\be}} &=-b_{\be\al} \, ,\\
       J^{F_{+}^{\al\be}} &=k\del c_{\be\al}-c_{\be\ga}J^{E_{+}^{\al\ga}}_B-c_{\ga\al}J^{E_{-}^{\ga\be}}_B-b_{\ga\de}c_{\be\ga}c_{\de\al}\, ,
    \end{split}
\end{align}
where the 
$J^{E_{\ep}^{\al\be}}_B$ are $\SLA{gl}{n}$ currents of level $\ep k-n$.
In the critical level case $k=0$, we can simply set the current corresponding to the central element of $\SLSA{gl}{n}{n}$
to zero
\begin{equation}
\sum_{\al}J^{E_{+}^{\al\al}}_B+J^{E_{-}^{\al\al}}_B\ = \ 0
\end{equation}
 and obtain the OPE relations of critical level $\SLSA{pgl}{n}{n}$. 
It is straightforward to check that this procedure of setting the central element to zero only works in the critical level limit.

\section{The real form $\SLSA{pu}{2,2}{4}$}\label{app:real}

We define the following automorphism $\omega$ of the complex Lie superalgebra $\SLSA{pgl}{4}{4}$.
\begin{equation}
\begin{split}
\omega(E^+_{\al\be})\ &= \ -E^+_{\be\al} \qquad \qquad \, \text{for} \ 1\leq\al\leq 2 \ \text{and} \ 1\leq\be\leq 2 \\
\omega(E^+_{\al\be})\ &= \ -E^+_{\be\al} \qquad \qquad \, \text{for} \ 3\leq\al\leq 4 \ \text{and} \ 3\leq\be\leq 4 \\
\omega(E^+_{\al\be})\ &= \ E^+_{\be\al} \qquad \qquad \ \ \,\, \text{for} \ 1\leq\al\leq 2 \ \text{and} \ 3\leq\be\leq 4 \\
\omega(E^+_{\al\be})\ &= \ E^+_{\be\al} \qquad \qquad \ \ \,\, \text{for} \ 3\leq\al\leq 4 \ \text{and} \ 1\leq\be\leq 2 \\
\omega(E^-_{\al\be})\ &= \ -E^-_{\be\al}  \\
\omega(F^+_{\al\be})\ &= \ -iF^-_{\be\al} \qquad \qquad \text{for}\  1\leq\al\leq 2  \\
\omega(F^+_{\al\be})\ &= \ iF^-_{\be\al} \ \ \qquad \qquad \, \text{for} \ 3\leq\al\leq 4  \\
\omega(F^-_{\al\be})\ &= \ -iF^+_{\be\al} \qquad \qquad \text{for}\  1\leq\be\leq 2  \\
\omega(F^-_{\al\be})\ &= \ iF^+_{\be\al} \qquad \qquad \ \ \, \text{for} \ 3\leq\be\leq 4  \\
\end{split}
\end{equation}
$\omega$ is not involutive but of order four. The semimorphism $\bar\omega$ defined as the concatenation of $\omega$ with complex conjugation
is involutive. 
The subalgebra of the complex Lie superalgebra $\SLSA{pgl}{4}{4}$ invariant under $\bar\omega$ is the real form $\SLSA{pu}{2,2}{4}$.
We can explicitly write a generating set over the real numbers. Let $X\in\{E^\pm_{\al\be},F^\pm_{\al\be}\}$ be the generators of $\SLSA{pgl}{4}{4}$, then
the elements $X+\omega(X)$ and $i(X-\omega(X))$ generate the real Lie superalgebra $\SLSA{pu}{2,2}{4}$ as a real superalgebra. 

In the literature, often other real form $\SLSA{pu}{2,2}{4}$ is considered. For example \cite{Dobrev:1985qz} uses the following semimorphism to define $\SLSA{pu}{2,2}{4}$ as invariant subalgebra of $\SLSA{sl}{4}{4}$
\begin{equation}
X_{(a)}\rightarrow -{(i)^a}\,\omega^{-1}X_{(a)}^{\dagger}\omega \quad,\quad\quad X_{(a)}\in\g_{a}
\end{equation}
where
$$
\newcommand*{\temp}{\multicolumn{1}{r|}{}} 
\omega=\left[\begin{array}{cccc} 0 &1_{2} &  \temp & 0\\  1_{2}&0 &\temp&0\\ \cline{1-4} 0& 0& \temp & 1_{N}
\end{array}\right] 
$$
This semimorphism is conjugate to our earlier definition by  the similarity matrix
$$
\newcommand*{\temp}{\multicolumn{1}{r|}{}} 
T=\left[\begin{array}{cccccc} 0 &1& 0 &-1 &  \temp & 0\\  1 &0&-1 &0 &  \temp & 0\\ 0 &1& 0 &1 &  \temp & 0\\ 1 &0& 1 &0 &  \temp & 0\\ \cline{1-6} 0& 0&0&0& \temp & 1_{N}
\end{array}\right] 
$$

Comparing the super-matrix presentation with expressions in Appendix D of \cite{Beisert:2004ry} (see also \cite{Kinney:2005ej}), we find the following identifications for our representation of the $\SLSA{pu}{2,2}{4}$ generators
$$
\newcommand*{\temp}{\multicolumn{1}{r|}{}} 
E=\left[\begin{array}{cccc} J_1{}_{\alpha}^{\beta} +{1\over2}\delta_{\alpha}^{\beta}D&P^{\alpha\dot\beta} &  \temp & Q^{\alpha\, b}\\  -K_{\dot\alpha\beta}&J_2{}_{\dot\beta}^{\dot\alpha} -{1\over2}\delta_{\dot\beta}^{\dot\alpha}D &\temp& -\bar S_{\dot\alpha}^b\\ \cline{1-4} S _a^{\beta}& \bar Q_{a\,\dot\beta}& \temp & R_{ab}
\end{array}\right] 
$$
Among the above generators we have also the following hermitian conjugation relations
$$
J_1^{\dagger}=-J_2\quad ,\quad K=P^{\dagger}
$$ 
These follow from the projection onto invariant subalgebra we are imposing. 
 The following identification also follows from the semimorphism 
$$
\bar Q=-i Q^{\dagger}\quad ,\quad \bar S=iS^{\dagger}
$$

\section{Conventions on weights}\label{app:weights}

We choose a basis of the Cartan subalgebra as follows
\begin{equation}
\begin{split}
H_1^+\ &= \ E^+_{11}-E^+_{22} \ \ , \ \ 
H_2^+\ = \ E^+_{22}-E^+_{33} \ \ , \ \ 
H_3^+\ = \ E^+_{33}-E^+_{44} \ \ , \\ 
H_1^-\ &= \ E^-_{44}-E^-_{33} \ \ , \ \ 
H_2^-\ = \ E^-_{33}-E^-_{22} \ \ , \ \ 
H_3^-\ = \ E^-_{22}-E^-_{22} \ \ . \ \
\end{split}
\end{equation}
Denote by $\lambda^\pm_i$ the dual linear form of $H^\pm_i$. 
Then roots and weights can be expressed in terms of these fundamental weights.
For example, the Weyl vector is
\begin{equation}
\rho \ = \ \lambda^+_1+\lambda^+_2+\lambda^+_3+\lambda^-_1+\lambda^-_2+\lambda^-_3\, .
\end{equation}
Let $A$ be the Cartan matrix of $\SLA{sl}{4}$ and decompose a weight as $\mu=\mu^++\mu_-$. 
Then the bilinear form on the weight space is given by 
\begin{equation}
B(\mu,\nu)\ = \ (\mu^+,A^{-1}\nu^+)- (\mu^-,A^{-1}\nu^-)
\end{equation}
and $(\ \ ,\ \ )$ denotes the standard inner product in $\mathbb C^3$.
We denote highest-weight representations by $(a_1,a_2,a_3;b_1,b_2,b_3)$, for a weight
\begin{equation}
\mu\ = \ (a_1-a_2)\lambda^+_1+ (a_2-a_3)\lambda^+_2+ (a_1+a_2+2a_3)\lambda^+_3+(b_1-b_2)\lambda^-_1+ (b_2-b_3)\lambda^-_2+ (b_1+b_2+2b_3)\lambda^-_3\, .
\end{equation}
Let us list all fermionic positive roots. They are all isotropic.
\begin{equation}
\begin{split}
\alpha^1_{11}\ &= \ \lambda_1^+-\lambda_3^- \ , \qquad \qquad \qquad 
\alpha^1_{12}\ = \ \lambda_1^++\lambda_3^--\lambda_2^- \ , \\ 
\alpha^1_{13}\ &= \ \lambda_1^+-\lambda_1^-+\lambda_2^- \ , \qquad\qquad   
\alpha^1_{14}\ = \ \lambda_1^++\lambda_1^-\ , \\ 
\alpha^1_{21}\ &= \ \lambda_2^+-\lambda^+_1-\lambda_3^- \ , \qquad\qquad   
\alpha^1_{22}\ = \ \lambda_2^+-\lambda^+_1+\lambda_3^--\lambda_2^- \ , \\ 
\alpha^1_{23}\ &= \ \lambda_2^+-\lambda^+_1-\lambda_1^-+\lambda_2^- \ , \qquad  
\alpha^1_{24}\ = \ \lambda_2^+-\lambda^+_1+\lambda_1^-\ , \\ 
\alpha^1_{31}\ &= \ \lambda_3^+-\lambda^+_2-\lambda_3^- \ , \qquad\qquad   
\alpha^1_{32}\ = \ \lambda_3^+-\lambda^+_2+\lambda_3^--\lambda_2^- \ , \\ 
\alpha^1_{33}\ &= \ \lambda_3^+-\lambda^+_2-\lambda_1^-+\lambda_2^- \ , \qquad 
\alpha^1_{34}\ = \ \lambda_3^+-\lambda^+_2+\lambda_1^-\ , \\ 
\alpha^1_{41}\ &= \ -\lambda_3^+-\lambda_3^- \ , \qquad\qquad   
\alpha^1_{42}\ = \ -\lambda_3^++\lambda_3^--\lambda_2^- \ , \\ 
\alpha^1_{43}\ &= \ -\lambda_3^+-\lambda_1^-+\lambda_2^- \ , \qquad  
\alpha^1_{44}\ = \ -\lambda_3^++\lambda_1^-\ , \\ 
\end{split}
\end{equation}
Let $\mu=\mu^++\mu^-$ such that the components satisfy $\mu^+_i=\mu^-_i$, then clearly
\begin{equation}\label{eq:atyp}
B(\mu+\rho,\alpha^1_{14}) \ = \ B(\mu+\rho,\alpha^1_{23}) \ = \ B(\mu+\rho,\alpha^1_{32}) \ = \ B(\mu+\rho,\alpha^1_{41}) \ = \ 0\,.
\end{equation}

\section{The Zhu Algebra}\label{app:zhu}

The Zhu algebra of the free theory is the polynomial algebra consisting of the rank $N^2$ Weyl algebra tensored with the rank $N^2$ Clifford algebra. 
These are generated by even coordinates $x=x_{ij}$ ($x$ a $N\times N$ matrix, $x_{ij}$ the components), their derivatives $d/dx=d/dx_{ij}$, odd coordinates $\theta=\theta_{ij}$ and their derivatives $d/d\theta=d/d\theta_{ij}$.
The map from free theory to Zhu algebra is:
\begin{equation}
\begin{split}
\beta_-\mapsto x\quad, \quad \beta_+\mapsto d/dx \quad,\quad \chi^L_-\mapsto \theta\quad,\quad \del\chi_+\mapsto d/d\theta\, .
\end{split}
\end{equation}
Fields $\phi(\beta_\pm, \chi^L_\pm)$ in the free theory are mapped to polynomials
$f(x,\theta,d/dx,d/d\theta)$ in the Zhu algebra. The map is given by the following description:

\subsubsection*{The Zhu functor}
Let $\cV$ be a vertex algebra with weight grading $\cV = \bigoplus_{n\in\mathbb{Z}} \cV_n$. The Zhu functor \cite{Zh} attaches to $\cV$ an associative algebra $A(\cV)$, together with a surjective linear map $\pi_{\text{zhu}}:\cV\ra A(\cV)$. For $a\in \cV_m$, and $b\in\cV$, define $$a*b = Res_z \bigg (a(z) \frac{(z+1)^{m}}{z}b\bigg),$$ and extend $*$ by linearity to a bilinear operation $\cV\otimes \cV\ra \cV$. Let $O(\cV)$ denote the subspace of $\cV$ spanned by elements of the form $$a\circ b = Res_z \bigg (a(z) \frac{(z+1)^{m}}{z^2}b\bigg),$$ for $a\in\cV_m$, and let $A(\cV)$ be the quotient $\cV/O(\cV)$, with projection $\pi_{\text{zhu}}:\cV\ra A(\cV)$. For $a,b\in \cV$, $a\sim b$ means $a-b\in O(\cV)$, and $[a]$ denotes the image of $a$ in $A(\cV)$. 
\begin{thm} (Zhu) $O(V)$ is a two-sided ideal in $V$ under the product $*$, and $(A(V),*)$ is an associative algebra with unit $[1]$. The assignment $\cV\mapsto A(\cV)$ is functorial.\end{thm}
Let $\mathcal{V}$ be a vertex algebra which is strongly generated by a set of weight-homogeneous elements $\alpha_i$ of weights $w_i$, for $i$ in some index set $I$. Then $A(\mathcal{V})$ is generated by $\{ a_i = \pi_{\text{zhu}}(\alpha_i(z))|~i\in I\}$. Moreover, $A(\mathcal{V})$ inherits a filtration (but not a grading) by weight, and the associated graded object $gr(A(\mathcal{V}))$ is a commutative algebra with generators $\{\bar{a}_i|~i\in I\}$.


\begin{thebibliography}{99}



\bibitem{Maldacena:1997re}
  J.~M.~Maldacena,
  Adv.\ Theor.\ Math.\ Phys.\  {\bf 2}, 231 (1998)
  [Int.\ J.\ Theor.\ Phys.\  {\bf 38}, 1113 (1999)]
  [arXiv:hep-th/9711200];
  S.~S.~Gubser, I.~R.~Klebanov and A.~M.~Polyakov,
  Phys.\ Lett.\  B {\bf 428}, 105 (1998)
  [arXiv:hep-th/9802109];
  E.~Witten,
  Adv.\ Theor.\ Math.\ Phys.\  {\bf 2}, 253 (1998)
  [arXiv:hep-th/9802150].
  
\bibitem{Konstein:2000bi}
  S.~E.~Konstein, M.~A.~Vasiliev and V.~N.~Zaikin,
  JHEP {\bf 0012}, 018 (2000)
  [arXiv:hep-th/0010239].
  
  
\bibitem{Sundborg}
P. Haggi-Mani and B. Sundborg,
``Free Large $N$ Supersymmetric Yang-Mills Theory as a String Theory,''
{\tt hep-th/0002189}; B. Sundborg,
``Stringy Gravity, Interacting Tensionless Strings and Massless
   Higher Spins,''
{\tt hep-th/0103247}.

\bibitem{EWtalk}
E. Witten, Talk at the John Schwarz 60-th Birthday Symposium, {\tt
http://theory.caltech.edu/jhs60/witten/1.html}.
  

\bibitem{Shaynkman:2001ip}
  O.~V.~Shaynkman and M.~A.~Vasiliev,
  Theor.\ Math.\ Phys.\  {\bf 128}, 1155 (2001)
  [Teor.\ Mat.\ Fiz.\  {\bf 128}, 378 (2001)]
  [arXiv:hep-th/0103208].


\bibitem{Sezgin:2001zs}
  E.~Sezgin and P.~Sundell,
  JHEP {\bf 0109}, 036 (2001)
  [arXiv:hep-th/0105001].
  
\bibitem{Vasiliev:2001zy}
  M.~A.~Vasiliev,
  Phys.\ Rev.\  D {\bf 66}, 066006 (2002)
  [arXiv:hep-th/0106149].

\bibitem{Polyakov:2001af}
  A.~M.~Polyakov,
  Int.\ J.\ Mod.\ Phys.\  A {\bf 17S1}, 119 (2002)
  [arXiv:hep-th/0110196].

\bibitem{Mikhailov:2002bp}
  A.~Mikhailov,
  arXiv:hep-th/0201019.
  
\bibitem{Sezgin:2002rt}
  E.~Sezgin and P.~Sundell,
  Nucl.\ Phys.\  B {\bf 644}, 303 (2002)
  [Erratum-ibid.\  B {\bf 660}, 403 (2003)]
  [arXiv:hep-th/0205131].


  
\bibitem{Isberg:1992ia}
  J.~Isberg, U.~Lindstrom and B.~Sundborg,
  Phys.\ Lett.\  B {\bf 293}, 321 (1992)
  [arXiv:hep-th/9207005];
  J.~Isberg, U.~Lindstrom, B.~Sundborg and G.~Theodoridis,
  Nucl.\ Phys.\  B {\bf 411}, 122 (1994)
  [arXiv:hep-th/9307108].
  
\bibitem{Segal:2002gd}
  A.~Y.~Segal,
  Nucl.\ Phys.\  B {\bf 664}, 59 (2003)
  [arXiv:hep-th/0207212].
  
\bibitem{Bianchi:2003wx}
  M.~Bianchi, J.~F.~Morales and H.~Samtleben,
  JHEP {\bf 0307}, 062 (2003)
  [arXiv:hep-th/0305052].


\bibitem{Lindstrom:2003mg}
  U.~Lindstrom and M.~Zabzine,
  Phys.\ Lett.\  B {\bf 584}, 178 (2004)
  [arXiv:hep-th/0305098].

\bibitem{Bonelli:2003zu}
  G.~Bonelli,
  JHEP {\bf 0311}, 028 (2003)
  [arXiv:hep-th/0309222].
  
\bibitem{Sagnotti:2003qa}
  A.~Sagnotti and M.~Tsulaia,
  Nucl.\ Phys.\  B {\bf 682}, 83 (2004)
  [arXiv:hep-th/0311257].


\bibitem{Bakas:2004jq}
  I.~Bakas and C.~Sourdis,
  JHEP {\bf 0406}, 049 (2004)
  [arXiv:hep-th/0403165].


\bibitem{Vasiliev:2001wa}
  M.~A.~Vasiliev,
  Nucl.\ Phys.\  B {\bf 616}, 106 (2001)
  [Erratum-ibid.\  B {\bf 652}, 407 (2003)]
  [arXiv:hep-th/0106200];
  X.~Bekaert, S.~Cnockaert, C.~Iazeolla and M.~A.~Vasiliev,
  arXiv:hep-th/0503128.






\bibitem{Klebanov:2002ja}
  I.~R.~Klebanov and A.~M.~Polyakov,
  Phys.\ Lett.\  B {\bf 550}, 213 (2002)
  [arXiv:hep-th/0210114].
  
\bibitem{Fradkin:1986qy}
  E.~S.~Fradkin and M.~A.~Vasiliev,
  Nucl.\ Phys.\  B {\bf 291}, 141 (1987);
  E.~S.~Fradkin and M.~A.~Vasiliev,
  Phys.\ Lett.\  B {\bf 189}, 89 (1987);
  M.~A.~Vasiliev,
  Phys.\ Lett.\  B {\bf 243}, 378 (1990);
  M.~A.~Vasiliev,
  Phys.\ Lett.\  B {\bf 567}, 139 (2003)
  [arXiv:hep-th/0304049].

  
  
\bibitem{Giombi:2009wh}
  S.~Giombi and X.~Yin,
  JHEP {\bf 1009}, 115 (2010)
  [arXiv:0912.3462 [hep-th]];
  S.~Giombi and X.~Yin,
  JHEP {\bf 1104}, 086 (2011)
  [arXiv:1004.3736 [hep-th]];
\bibitem{Giombi:2011ya}
  S.~Giombi and X.~Yin,
  arXiv:1105.4011 [hep-th].

\bibitem{Douglas:2010rc}
  M.~R.~Douglas, L.~Mazzucato and S.~S.~Razamat,
  Phys.\ Rev.\  D {\bf 83}, 071701 (2011)
  [arXiv:1011.4926 [hep-th]].




\bibitem{Berkovits:2007zk}
  N.~Berkovits,
  JHEP {\bf 0708 } (2007)  011.
  [hep-th/0703282].

\bibitem{Berkovits:2007rj}
  N.~Berkovits, C.~Vafa,
  JHEP {\bf 0803 } (2008)  031.
  [arXiv:0711.1799 [hep-th]].
  
\bibitem{Berkovits:2008qc}
  N.~Berkovits,
  JHEP {\bf 0809}, 088 (2008).
  [arXiv:0806.1960 [hep-th]].


\bibitem{Berkovits:2008ga}
  N.~Berkovits,
  JHEP {\bf 0909}, 051 (2009).
  [arXiv:0812.5074 [hep-th]].


\bibitem{Berkovits:2003pq}
  N.~Berkovits, H.~Ooguri, C.~Vafa,
  Commun.\ Math.\ Phys.\  {\bf 252}, 259-274 (2004).
  [hep-th/0310118].

  \bibitem{CGL}
  T.~Creutzig, P.~Gao, A.~R.~Linshaw, 
  [arXiv:1109.4065 [math.QA]].

\bibitem{Zamolodchikov:1985wn}
  A.~B.~Zamolodchikov,
  Theor.\ Math.\ Phys.\  {\bf 65}, 1205 (1985)
  [Teor.\ Mat.\ Fiz.\  {\bf 65}, 347 (1985)].

\bibitem{Feigin:2004wb}
  B.~L.~Feigin and A.~M.~Semikhatov,
  Nucl.\ Phys.\  B {\bf 698}, 409 (2004)
  [arXiv:math/0401164].

\bibitem{CR}
  T.~Creutzig, D.~Ridout,
  [arXiv:1111.5049 [hep-th]].

  \bibitem{Zh}
  Yongchang Zhu, 
  J. Amer. Math. Soc. 9 (1996), 237-302.



\bibitem{Rozansky:1992rx}
  L.~Rozansky, H.~Saleur,
  Nucl.\ Phys.\  {\bf B376 } (1992)  461-509.

\bibitem{Schomerus:2005bf}
  V.~Schomerus, H.~Saleur,
  Nucl.\ Phys.\  {\bf B734 } (2006)  221-245.
  [hep-th/0510032].

\bibitem{Saleur:2006tf}
  H.~Saleur, V.~Schomerus,
  Nucl.\ Phys.\  {\bf B775 } (2007)  312-340.
  [hep-th/0611147].

\bibitem{Gotz:2006qp}
  G.~Gotz, T.~Quella, V.~Schomerus,
  JHEP {\bf 0703 } (2007)  003.
  [hep-th/0610070].

\bibitem{Quella:2007hr}
  T.~Quella, V.~Schomerus,
  JHEP {\bf 0709 } (2007)  085.
  [arXiv:0706.0744 [hep-th]].

\bibitem{Creutzig:2007jy}
  T.~Creutzig, T.~Quella, V.~Schomerus,
  Nucl.\ Phys.\  {\bf B792 } (2008)  257-283.
  [arXiv:0708.0583 [hep-th]].

\bibitem{Creutzig:2009fh}
  T.~Creutzig, P.~B.~Ronne, V.~Schomerus,
  Phys.\ Rev.\  {\bf D80 } (2009)  066010.
  [arXiv:0907.3902 [hep-th]].

\bibitem{Hikida:2007sz}
  Y.~Hikida, V.~Schomerus,
  JHEP {\bf 0712 } (2007)  100.
  [arXiv:0711.0338 [hep-th]].


\bibitem{Creutzig:2009zz}
  T.~Creutzig,
  [arXiv:0908.1816 [hep-th]].

\bibitem{Creutzig:2008an}
  T.~Creutzig, P.~B.~Ronne,
  Nucl.\ Phys.\  {\bf B815 } (2009)  95-124.
  [arXiv:0812.2835 [hep-th]].

\bibitem{Creutzig:2008ag}
  T.~Creutzig,
  Nucl.\ Phys.\  {\bf B812 } (2009)  301-321.
  [arXiv:0809.0468 [hep-th]].

\bibitem{Creutzig:2008ek}
  T.~Creutzig, V.~Schomerus,
  Nucl.\ Phys.\  {\bf B807 } (2009)  471-494.
  [arXiv:0804.3469 [hep-th]].
  
\bibitem{Creutzig:2010zp}
  T.~Creutzig, Y.~Hikida,
  Nucl.\ Phys.\  {\bf B842 } (2011)  172-224.
  [arXiv:1004.1977 [hep-th]].

\bibitem{Creutzig:2011cu}
  T.~Creutzig, D.~Ridout,
  [arXiv:1107.2135 [hep-th]].

\bibitem{Creutzig:2011qm}
  T.~Creutzig, Y.~Hikida, P.~B.~Ronne,
  JHEP {\bf 1106 } (2011)  063.
  [arXiv:1103.5753 [hep-th]].



\bibitem{Quella:2007sg}
  T.~Quella, V.~Schomerus, T.~Creutzig,
  JHEP {\bf 0810 } (2008)  024.
  [arXiv:0712.3549 [hep-th]].

\bibitem{Candu:2009ep}
  C.~Candu, V.~Mitev, T.~Quella, H.~Saleur, V.~Schomerus,
  JHEP {\bf 1002 } (2010)  015.
  [arXiv:0908.0878 [hep-th]].

\bibitem{Mitev:2008yt}
  V.~Mitev, T.~Quella, V.~Schomerus,
  JHEP {\bf 0811 } (2008)  086.
  [arXiv:0809.1046 [hep-th]].


\bibitem{Konechny:2010nq}
  A.~Konechny, T.~Quella,
  JHEP {\bf 1103 } (2011)  124.
  [arXiv:1011.4813 [hep-th]].

\bibitem{Creutzig:2010hr}
  T.~Creutzig,
  Nucl.\ Phys.\  {\bf B849 } (2011)  636-653.
  [arXiv:1011.6424 [hep-th]].

\bibitem{Candu:2010yg}
  C.~Candu, T.~Creutzig, V.~Mitev, V.~Schomerus,
  JHEP {\bf 1005 } (2010)  047.
  [arXiv:1001.1344 [hep-th]].

\bibitem{Ashok:2009xx}
  S.~K.~Ashok, R.~Benichou, J.~Troost,
  JHEP {\bf 0906 } (2009)  017.
  [arXiv:0903.4277 [hep-th]].

\bibitem{Benichou:2010rk}
  R.~Benichou, J.~Troost,
  JHEP {\bf 1004 } (2010)  121.
  [arXiv:1002.3712 [hep-th]].



\bibitem{Creutzig:2010ne}
  T.~Creutzig, P.~B.~Ronne,
  JHEP {\bf 1011 } (2010)  021.
  [arXiv:1006.5874 [hep-th]].



\bibitem{Dobrev:1985qz}
  V.~K.~Dobrev and V.~B.~Petkova,
  Fortsch.\ Phys.\  {\bf 35}, 537 (1987).


\bibitem{Beisert:2004ry}
  N.~Beisert,
  Phys.\ Rept.\  {\bf 405}, 1 (2005)
  [arXiv:hep-th/0407277].

\bibitem{Kinney:2005ej}
  J.~Kinney, J.~M.~Maldacena, S.~Minwalla and S.~Raju,
  Commun.\ Math.\ Phys.\  {\bf 275}, 209 (2007)
  [arXiv:hep-th/0510251].
  
  

\end{thebibliography}
\end{document}